# Designing a Magnetic Measurement Data Acquisition and Control System with Reuse in Mind: A Rotating Coil System Example

J. M. Nogiec, P. Akella, G. Chlachidze,  J. DiMarco, M. Tartaglia, P. Thompson, K. Trombly-Freytag, and D. Walbridge

*Abstract*—**Accelerator magnet test facilities frequently need to measure different magnets on differently equipped test stands and with different instrumentation. Designing a modular and highly reusable system that combines flexibility built-in at the architectural level as well as on the component level addresses this need. Specification of the backbone of the system, with the interfaces and dataflow for software components and core hardware modules, serves as a basis for building such a system. The design process and implementation of an extensible magnetic measurement data acquisition and control system are described, including techniques for maximizing the reuse of software. The discussion is supported by showing the application of this methodology to constructing two dissimilar systems for rotating coil measurements, both based on the same architecture and sharing core hardware modules and many software components. The first system is for production testing 10 m long cryo-assemblies containing two MQXFA quadrupole magnets for the high-luminosity upgrade of the Large Hadron Collider and the second for testing IQC conventional quadrupole magnets in support of the accelerator system at Fermilab.**

## I. Introduction

THE Fermilab Magnet Test Facility tests various types of accelerator magnets, including  permanent, normal conducting (resistive) and  superconducting magnets.  The tests are conducted on horizontal and vertical test stands. Various measurement methods, probes, and DAQ and instrumentation hardware are used in these systems. Despite these variations, there are common functionalities and similarities that allow for a shared architecture and software reuse.

To efficiently support diverse test efforts at the Fermilab Magnet Test Facility (MTF), a solution allowing for a high level of reuse and significant software flexibility was implemented. This solution adopts the product line method to develop a family of magnet measurement systems, which reduces the cost of developing new systems and  allows for configuration of existing systems to  support variants of measurements.

The application of product line development at Fermilab follows the latest trends in software engineering for addressing reuse. The earlier approaches to provide a uniform platform for developing magnetic measurement systems include the use of checklists to describe measurement sequences [1], graphically configurable component-based systems [2], and a domain-specific language to describe magnetic measurements [3].

The product line approach is outlined and supported by presenting implementations of two different rotating coil magnet measurement systems, both of which are part of the same product family.

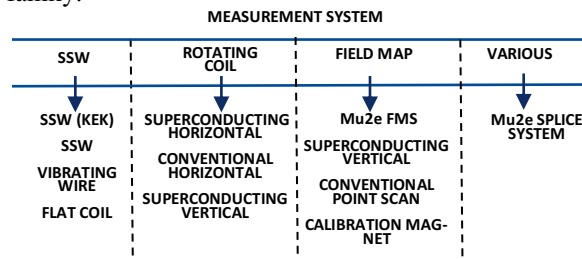

Fig. 1.  Magnet measurement systems family tree.

## II. Family of Magnetic Measurement Systems

### A.  Magnet Measurement Systems Family Tree

The measurement systems being used at MTF can be classified into four branches based on the dominant measurement method (Fig. 1):
1) stretched wire systems, including single-wire systems (SSW) with various modes of operation (AC, DC, vibrating wire) and flat coil systems;
2) rotating coil systems, including systems with various types of coils and positioning systems;
3) field-mapping systems, including point scan systems and 3D mapping systems, using Hall and NMR probes; and
4) miscellaneous test systems, specialized systems that cannot be classified as belonging to the other branches.

### B.  Variability in Magnetic Measurement Systems

In order to develop a  family of measurement systems, one needs to analyze the differences and similarities between various systems, referred to as variability. The similarities allow for designing a common architecture and software artifacts

This work was supported in part by the Fermi Research Alliance under DOE Contract DE-AC02-07CH11359. *(Corresponding author: Jerzy Nogiec.)*

J. Nogiec P. Akella, G. Chlachidze, J. DiMarco, M. Tartaglia, P. Thompson, K. Trombly-Freytag and D. Walbridge are with Fermi National Accelerator Laboratory, Batavia, IL 60510 USA (e-mail: nogiec@fnal.gov).

providing reusable functionality. Differences between various test system family members stem from different measurement methods and their implementations. The differences are categorized into one of the following areas:

1) hardware organization, with different data acquisition modules and instruments, sensors and probes, positioning and motion systems, and power supplies and their control systems;
2) software functionality, which stems from choices of hardware, measurement procedures, data visualization, on-line analysis and quality checks, and data models for processing, archival and calibrations;
3) integration with other systems, such as cryogenic control systems, electronic logbooks, external power supply systems, motion systems, and data services.

## III. DESIGNING FOR REUSE

Though magnetic measurement systems may have different hardware and use different measurement methods, there is a similarity in their general functionality and architecture.

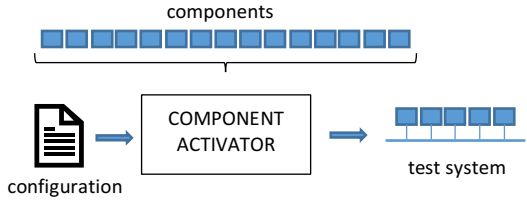

Fig. 2. Dynamically assembling test systems from reusable components.

### A. Software Product Line Development

Software product line development discussed in [4] is a methodology used to achieve reuse when developing a set of related applications in a particular application domain. This methodology has been adopted at the Fermilab MTF to develop a family of measurement systems. A common architecture was used, which allows for configuring specific measurement systems from a set of reusable components (Fig. 2).

### B. Measurement Software Framework

The authors have combined the aforementioned features with scripting to coordinate the work of components and implement various measurement algorithms [5]. This paradigm has been used to develop a flexible measurement software framework, which is a common foundation for all developed test systems [6]. The framework provides a set of system components and a software bus for inter-component communication.

All components are based on the same template, utilizing a state machine and connectors to the software bus. Each component can be tailored via its configurable properties. Some components are designed to be universal and can accommodate plug-ins to alter their functionality.

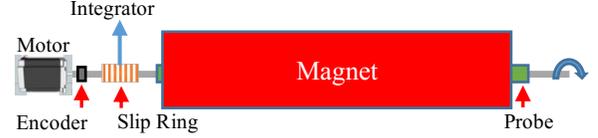

Fig. 3. Principles of a rotating coil system.

## IV. PRINCIPLES OF A ROTATING COIL SYSTEM

A coil having one or more windings is inserted in the opening of the accelerator magnet and rotated by a motor in the magnetic field (Fig. 3). Each winding, a number of wire loops, measures the changes of the intercepted flux. A digital integrator externally triggered by an angular encoder measures the differences in flux, $\Delta\Phi$, between two incremental angular positions, expressed as

$$\Delta\Phi = \Phi_{n+1} - \Phi_n = -\int_{t_n}^{t_{n+1}} V(t) \cdot dt. \qquad (1)$$

The field in the accelerator magnet can be described in terms of a harmonic expansion, with field quality expressed as the magnitude of undesirable harmonic terms in this expansion. Typical steps of the harmonic analysis include the following:

1) Accumulate $\Delta\Phi$ into $\Phi$ by performing a running sum of the flux data for each integrator channel, resulting in flux waveforms. Accumulate delta time in a similar manner.
2) Remove integrator drift from the waveforms. Any offset in the signal will be integrated and will result in drift. It is assumed that the drift is linear during one probe rotation. Drift as a function of time is subtracted from the waveform.
3) Perform either an FFT or curve fitting on the waveforms to find the Fourier amplitudes and phase angles, $\chi_n$, of the harmonic components. Use probe sensitivity factors on the Fourier amplitudes to obtain the harmonic amplitudes $C_n$.
4) Calculate the reference amplitude for the dominant multipole ($C_2$, where n = 2 for a quadrupole).
5) Calculate normalized harmonics coefficients, $c_n$, for quadrupole magnets as

$$c_n = (C_n/C_2) \cdot r^{n-2}, \qquad (2)$$

where r is the reference radius. The reference radius can be an arbitrary choice, but often it is chosen to be the probe radius or some standard length. Fermilab typically uses 25.4 mm (one inch) as a reference radius.
6) To standardize values in units, calculate normal and skew harmonics $b_n$ and $a_n$ from the normalized harmonics by using

$$b_n + i\,a_n = 10{,}000 \cdot c_n \cdot (\cos(\chi_n) + i\sin(\chi_n))\ . \qquad (3)$$

7) Determine the probe offsets from the magnetic center and use them to recalculate harmonics for centering correction.

For more comprehensive explanation of the theory of magnetic measurements with rotating coils, refer to [7] and [8].

## V. Magnet Measurements with Rotating Coils

To demonstrate the reuse effectiveness of product line development, two contrasting rotating system setups for testing a large superconducting magnet and a smaller resistive magnet are presented.

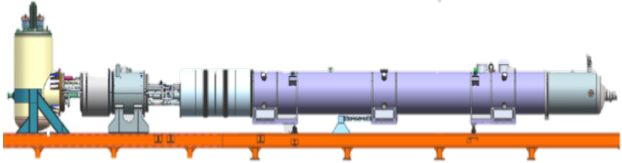

Fig. 4. Magnet assembly on a cryogenic test stand.

### A. Rotating Coil System for Superconducting Magnets

Fermilab will fabricate large aperture interaction region quadrupoles for final beam focusing as part of the High Luminosity LHC Accelerator Upgrade Project. LQXFA, a 10.1 m long cryogenic assembly, contains two 4.5 m $Nb_3Sn$ MQXFA magnets. At a nominal current of 16.23 kA, they produce a field gradient of 132.9 T/m.

The tests of the LQXFA magnets will be done on the Fermilab horizontal cryogenic test stand (Fig. 4). There will be two types of measurements:

1) a Z-scan measurement, where the probe is positioned at 109 mm increments starting outside of the magnetic field and stepping through the length of the magnet until it extends beyond the field at the opposite end, and
2) a current profile measurement, where current is raised and lowered according to a provided profile with the probe stationary.

The Z-scan will be repeated at multiple currents. The profile will be done at least at two positions, one in the body field of each MQXFA magnet.

A dedicated probe has been developed to test the MQXFA magnets. The system uses a short mole-type probe with three coils (436mm and 2 x 109 mm), each having three windings: unbucked, dipole-bucked, and dipole-quadrupole-bucked [9].

The combination of probes in the mole will allow for the measurement of the integrated quadrupole field and integrated harmonics (up to n = 14). In addition, the probe has been constructed so as to determine magnetic length, field direction, magnet center separation, end field harmonics, local twist variation of the roll angle, and the waviness of the local magnetic axis.

A multi-channel extensible integrator developed at Fermilab [10] acquires signals from the probe's windings, and its auxiliary digital and analog inputs allow it to receive a real-time stream of current readouts and to read the probe accelerometers.

The axial movement of the probe is accomplished with an approximately 10 m long translation table, controlled by a Galil controller. Probe rotation is supplied by a stepping motor controlled by an Aerotech controller. The probe position and its orientation are measured by a Leica AT960 laser tracker fully integrated with the measurement system.

The measurement system is integrated with the iFIX cryogenic control system to monitor magnet temperature and the power supply control system to control and monitor current. It is also integrated with an electronic logbook for making automated entries.

### B. Rotating Coil System for Normal Conducting Magnets

The IQC quadrupole magnets, along with IQB's and IQD's, are the primary focusing and defocusing magnets used in the Main Injector accelerator at Fermilab [11]. 2.5 m long IQC magnets produce a gradient of 19.6 T/m at the nominal current of 3.63 kA.

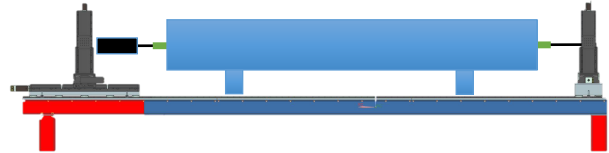

Fig. 5. Magnet on a room-temperature test stand.

The tests of the IQC magnets will be performed on a room-temperature stand equipped with water cooling (Fig. 5). The Morgan probe is positioned in the opening of the magnet with two Aerotech stages, one 3D and one 2D, supporting it in this position. The probe extends beyond the ends of the magnet, so only one measurement position is required.

The tests will be comprised of strength and harmonics measurements. Measurements will ramp the current to a maximum of 4 kA then back to 0 A, taking data at each step.

The data from several Morgan coil windings (2P, 4P, 6P, 8P, 10P, 12P, and 20P) will be acquired simultaneously by the Fermilab extensible integrator. Integrated strength will be obtained through analysis of the data from the quadrupole coil. Harmonics results will come from analysis of the data from the remaining windings, with harmonics up to 20-pole (n = 10) to be provided. Most of the harmonics are expected to be +/- several units or less, but the range of expected values varies from harmonic to harmonic.

The measurement system is integrated with the power supply control system to control and monitor current and to an electronic logbook for making automated entries.

## VI. Reuse of Components in Rotating Coil Systems

### A. Coordination

Different measurement procedures and specific scenarios are orchestrated by the same Script component, but with different Python coordination scripts and parameters.

### B. Devices and Instrumentation

Hardware standardization allows for reduced effort when integrating devices into the test systems. The same integrator component is used to acquire data from diametrically different probes in both systems, with its properties providing information about the signals on each channel.

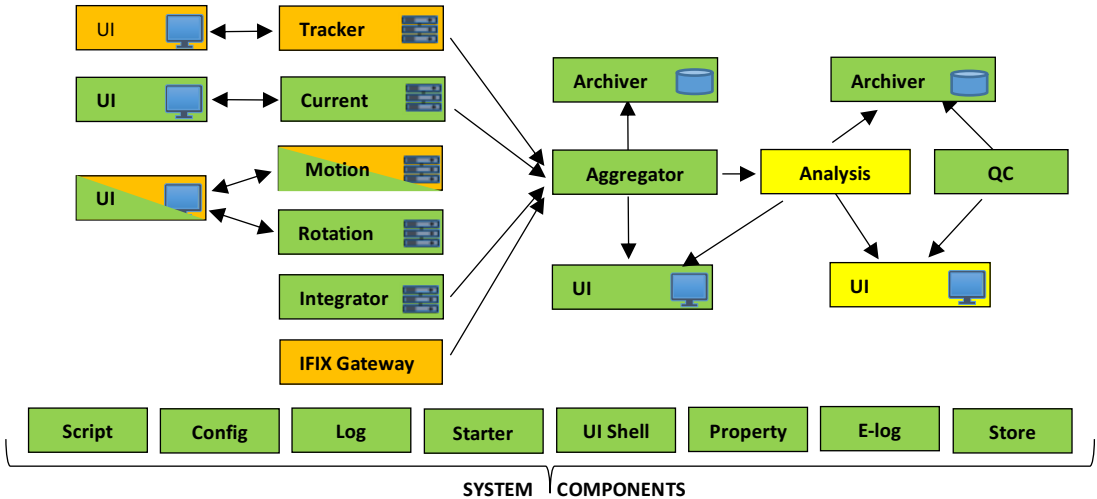

Fig. 6.  Software components and their reuse: complete reuse (green), partial reuse (yellow), and no reuse (orange).

The probe rotation system is using the same Aerotech controller, so the same software component can be used in both systems.

The probe positioning mechanism is different between the two systems.. The cryogenic test stand uses a long translation table, which is controlled by a dedicated component. The test stand for conventional magnets uses the Aerotech stages; therefore, can reuse the component for controlling Aerotech motion hardware developed for the stretched wire system [5].

The laser tracker is only used with the long superconducting magnet, but the Leica laser tracker is also used in a field mapping system [12].

### C.  Data Processing

Both systems use the same universal analysis component, but with different plug-ins due to differences in the analysis algorithms. The QC component has a slightly different set of quality check expressions specified in its properties for each system.

### D.  User Interface

The systems use the same Shell system component to host their user interface plug-ins. The flux waveforms and harmonics are shown using the same plug-in, but additional plug-ins display results of the Z-scan and the profile tests as functions of probe position and time, respectively.

### E.  Data Persistence

The data persistence and archival components are the same, due to their ability to save any data structure.

### F.  Integration with Other Systems

Both systems are integrated with their power supply control systems using the same component interacting with the universal power supply controllers.  Should interfacing with a different power supply control be needed, the component can be configured with a different plug-in. Superconducting magnets also require integration with the cryogenic control system, which is not needed in the case of room temperature magnets.

### G.  Components and Dataflow

The described rotating coil systems have similar data flow and share many components (Fig. 6). Three levels of component reuse are identified: a) complete reuse, when the same component is used in both systems, b) partial reuse, when the same component is used with different plug-ins, and c) no reuse, when different components are used or the functionality is not needed.

## VII.  Conclusion

Engineers and physicists at the Fermilab MTF must satisfy a constant demand for various new test and measurement systems required for testing different types of magnets, which are often one of a kind.

Effective software reuse and built-in flexibility are needed for reduction in cost and development time, and improved software maintainability. In response to this demand, the product line method was adopted to development of a family of magnet measurement systems based on a common architecture and framework. This method, coupled with component-based development, led to a change of system development paradigm from programming to configuring and assembling components and tailoring their behavior. The results are flexible measurement systems characterized by a high level of reuse and an agile measurement facility capable of quickly producing different types and versions of measurements.

The presented rotating coil systems, although quite different, are still characterized by a high level of reuse, which confirms the effectiveness of the adopted methodology.  Besides their inherent flexibility, they offer some advanced features such as measurement automation, embedded support for hardware debugging, system self-monitoring, run-time data analysis and visualization, and automated data quality control.

The solution of using product line development and configurable component-based systems is unique to the Fermilab Magnet Test Facility and novel in the area of accelerator magnet test and measurement systems.